\documentclass[preprint]{aastex}
\usepackage{graphicx}
\usepackage{subfigure}
\usepackage{natbib}

\newcommand{\mycitet}[1]{\citet{#1}}
\newcommand{\mycitep}[1]{\citep{#1}}
\newcommand{\myeq}[1]{Eq. \ref{#1}}
\newcommand{\myfig}[1]{Fig. \ref{#1}}
\newcommand{\mytable}[1]{Table \ref{#1}}
\newcommand{\myfigs}[2]{Figs. \ref{#1}--\ref{#2}}

\title{Design of a telescope-occulter system for THEIA}
\author{Eric Cady$^{1}$, Ruslan Belikov$^{2}$, Philip Dumont$^{3}$, Robert Egerman$^{4}$, N. Jeremy Kasdin$^{1}$, \\ Roger Linfield$^{5}$, Doug Lisman$^{3}$, Dmitry Savransky$^{1}$, Sara Seager$^{6}$, Stuart Shaklan$^{3}$, \\ David Spergel$^{7}$, Domenick Tenerelli$^{8}$, and Robert Vanderbei$^{9}$}

\begin{abstract}
The Telescope for Habitable Exoplanets and Interstellar/Intergalactic Astronomy (THEIA) is a mission concept study for a flagship-class telescope-occulter system to search for terrestrial planets and perform general astrophysics with a space-based 4m telescope. A number of design options were considered for the occulter and telescope optical systems; in this paper we discuss the design of occulters and coronagraphs for THEIA and examine their merits. We present two optimized occulters: a $25.6$m-radius occulter with $19$m petals that achieves $10^{-12}$ suppression from $250-1000$nm with a $75$mas inner working angle, and a $20.0$m-radius occulter with $10$m petals that achieves $10^{-12}$ suppression from
$250-700$nm with a $75$mas inner working angle.  For more widely separated planets (IWA $> 108$mas), this second occulter is designed to operate at a second closer distance where it provides $10^{-12}$ suppression from $700-1000$nm.  We have also explored occulter/coronagraph hybrid systems, and found that an AIC coronagraph that exploits the symmetry of the PSF at the occulter can improve performance; however, it requires very accurate tolerances on the occulter manufacturing of the telescope/occulter system as the AIC does not can asymmetric terms.  Other coronagraphs proved infeasible, primarily due to the fact that the residual starlight from the occulter is not a plane wave, and so is poorly suppressed by the coronagraph.
\end{abstract}


\begin{document}
\maketitle
\footnotetext[1]{Dept. of Mechanical and Aerospace Engineering, Princeton University, Princeton, NJ 08544.  Contact author: ecady@princeton.edu}
\footnotetext[2]{NASA Ames Research Center, Moffett Field, Sunnyvale, CA 94088}
\footnotetext[3]{Jet Propulsion Laboratory, 4800 Oak Grove Drive, Pasadena, CA 91109}
\footnotetext[4]{ITT Space Systems Division, 800 Lee Road, Rochester, NY  14606}
\footnotetext[5]{Ball Aerospace, 1600 Commerce Street, Boulder, CO 80301}
\footnotetext[6]{Dept. of Physics, Massachusetts Institute of Technology, 77 Massachusetts Avenue, Cambridge, MA 02139}
\footnotetext[7]{Dept. of Astrophysical Sciences, Princeton University, Princeton, NJ 08544}
\footnotetext[8]{Lockheed Martin Space Systems, 1111 Lockheed Martin Way, Sunnyvale, CA 94089}
\footnotetext[9]{Dept. of Operations Research and Financial Engineering, Princeton University, Princeton, NJ 08544}
\pagebreak

\section{Introduction}

An external occulter, or \emph{occulter} for short, is an opaque screen that is placed in front of a telescope for high-contrast imaging.  Its purpose is to eliminate most of the light from a bright stellar object before it ever reaches the telescope, so that dim objects near the bright one are not overwhelmed by stray light from scattering and aberrations in the telescope optics.

Occulters were first proposed for solar coronagraphy in 1948 \mycitep{Eva48}, and for finding extrasolar planets in 1962 \mycitep{Spi62}. Occulters designed for solar coronagraphy, such as those flown on SOHO \mycitep{Bru95} and STEREO \mycitep{How00}, are generally located a few meters from the telescope optics, and are attached to the telescope, allowing the entire optical system to be on a single craft.  More recently, solar occulters have been proposed to fly detached from the spacecraft at a distance on the order of $100$m \mycitep{Viv07}. On the other hand, occulters designed for extrasolar planet searches, such as those proposed in (\mycitet{Cop00, Sch03, Cas06, Van07}), are not only detached from the spacecraft, but are located tens or hundreds of thousands of kilometers away.  This distance gives them a small angular size, theoretically allowing objects within $100$mas or less of a star to be imaged.

In this paper, we focus on designs for exoplanet imaging applications.  The difference in the analysis comes primarily from the size of the objects being occulted; the angular size of a star being examined for planets is a fraction of a milliarcsecond, and so the star is generally treated as a point source and the incident electric field as a plane wave.

While some occulter designs, such as BOSS \mycitep{Cop00}, use materials with spatially-varying transmittance, most proposed occulter designs are binary, either blocking the light entirely at a point or letting the light through.  This sidesteps a number of engineering difficulties in creating the transmitting occulter, such as wavelength-dependent transmittance.  It requires, however, that the binary occulter be made with edges following precisely manufactured curves that are designed so that the diffracted light in the shadow behind the occulter is nearly nulled out.  These curves are known as \emph{petals}, and a binary occulter takes on a flower-like shape.

Occulting systems can be separated further into pure and hybrid designs.  The pure occulter design achieves all of its light suppression from the occulter alone; a simple diffraction-limited telescope simply needs to be placed in the region of destructive interference to view objects around the target star.  This design has the advantage of eliminating problems from light scatter within the telescope, since the intensity of the starlight is already reduced below the intensity of the planet in the image plane.  A hybrid system adds a coronagraph in the telescope, to provide part of the light suppression.  The hybrid approach introduces additional complexity to the system, and potentially requires tighter tolerances to ensure the occulter and coronagraph work together.  On the other hand, it potentially allows the occulter to be built smaller and flown closer than a pure occulter, which reduces mass and slew time between targets.

One system studied which would use an occulter for planet-finding is the Telescope for Habitable Exoplanets and Interstellar/Intergalactic Astronomy (THEIA).  The THEIA mission consists of a 4m on-axis telescope and an occulter flown tens of thousands of kilometers out.  The telescope contains three instruments: an ultraviolet spectrograph (UVS), a wide-field imager called the Star Formation Camera (SFC), and an optical system for imaging planets in conjunction with an occulter: the eXtrasolar Planet Characterizer (XPC).  The UVS and SFC would perform general astrophysics while the occulter is slewing from target to target.  The telescope and occulter would be launched from separate Atlas V launch vehicles, and would be placed in the vicinity of a halo orbit about the Sun-Earth L2 point.  The telescope would remain on the halo orbit, while the occulter would be slewed to align between telescope and target.

In the process of finalizing a design for THEIA, both pure and hybrid occulter systems were considered.  In this paper, we will compare their performance, and present some designs that would achieve the performance required to see planets about a nearby star.  We find that hybrids do not perform as well as hoped, and the best designs are pure occulters operated at multiple distances from the telescope.

\section{Science goals and requirements}

The prime scientific goal of THEIA's XPC and occulter is to find and characterize Earth-like planets. By Earth-like we mean an Earth-size planet in an Earth-like orbit about a Sun-like star.  The overarching goal can be divided into discovery and characterization, which flow down into requirements.

\noindent\emph{Discovery:} THEIA is designed to be capable of finding Earth-size planets in Habitable Zone orbits \mycitep{Kas93} around nearby stars by direct imaging. (The value of $\eta_{\oplus}$, the expected number of Earth-mass planets per star, is unknown, though data from Kepler is expected to constrain this value \mycitep{Bas05}.)  While THEIA is designed to find small, rocky planets, it will detect all objects brighter than a terrestrial planet in the same system, including super-Earths (1-10 Earth-mass planets), gas giants, and brown dwarfs.  The variety of direct imaging planet discoveries will improve our knowledge of planet distribution within a planetary system and will help to constrain theories of planet formation.

\noindent\emph{Orbital parameters:}
Multiple observations will help to constrain the orbital elements $(a, e, i, \Omega, \omega, \nu)$ and the period of a planet, using observations of the planet's location, motion over time, and change in flux with time \mycitep{Bur05}.   Planetary masses may be estimated by using models of the planet's structure, calculations of the planetary period and semi-major axis, and cross-correlation with measurement from indirect detection methods.

\noindent\emph{Spectral/photometric characterization:} Spectral characterization can tell us about the contents of the atmosphere or the planet's surface. At THEIA wavelengths, H$_{2}$O has strong absorption features and for an Earth-like planet indicates surface liquid water. All life on Earth requires liquid water and so water vapor is an important spectral feature. Spectral characterization can also reveal the presence of biosignatures--- gases whose presence  indicates a chemical disequilibrium that could potentially be caused by biological mechanisms.  Oxygen and its photochemical byproduct ozone are Earth's most robust biosignatures. Methane is a possible biosignature from early Earth. \mycitep{Des02} Additionally, fluctuations in photometric intensity can be used to determine planet rotation rates, cloud cover, and potentially even latitudinally-averaged land distribution on a rocky planet \mycitep{For01, Pal08, Cow09}.

\noindent These scientific goals lead to the following requirements on our occulter/telescope system:
\begin{itemize}
\item The flux from an Earth-sized rocky planet 1AU from its parent star is approximately $10^{10}$ times fainter than the flux from the star itself, with some dependence due to albedo and planetary phase.  The angular separation of these objects is dependent on the distance to the star: at 10pc, THEIA must be able to discern an object $10^{10}$ times dimmer at 100 mas angular separation; at 20pc, this drops to 50mas.  From a scientific perspective, THEIA should be able to discern objects with as small an angular separation as possible.
\item Despite the uncertainty in $\eta_{\oplus}$, THEIA needs to be capable of detecting Earth-like planets even if their frequency is as low as 0.1.
\item To reliably detect the sharp $O_{3}$ band cutoff at 300 nm, THEIA needs to be able to take the spectrum of a planet down to 250nm.  To detect the $H_{2}O$ band at 940 nm, THEIA needs to be able to take the spectrum up to $1000$nm. THEIA should also be capable of spectral characterization at all intermediate wavelengths.
\end{itemize}

\section{Occulter design}

\subsection{Propagation equations}

The electric field downstream of a plane wave of unit intensity normally incident on an occulter can be written in polar coordinates as:
\begin{eqnarray}
    E_{\mathrm{occ}}(\rho, \phi) &=& E_0 e^{i k z} \left(1 - \frac{2 \pi}{i \lambda z} \int^R_0 A(r) J_0\left(\frac{2 \pi r \rho}{\lambda z}\right) e^{\frac{i k}{2 z}\left(r^2 + \rho^2\right)} r dr\right) \nonumber \\
    &&- E_0 e^{i k z}\sum^{\infty}_{j = 1}\frac{(-1)^j k}{i z}
    \left(\int^R_0 e^{\frac{i k}{2
    z}(r^2+\rho^2)} J_{j N}\left(\frac{2 \pi r \rho}{
    \lambda z}\right) \frac{\sin{(j \pi A(r))}}{j \pi} r dr\right) \nonumber \\
    &&\qquad \qquad
    \times \left(2\cos{(j N (\phi-\pi/2))}\right)\label{prop1}
\end{eqnarray}
where $N$ is the number of petals of the occulter and $A(r)$ is a function describing how the petal width changes with radius.  A circle drawn at radius $r$ will have an alternating series of petals and gaps; the fraction of the circumference of the circle that is covered by petal is $A(r)$.  (If the occulter were smoothly apodized rather than one-zero everywhere, then $A(r)$ would be the apodization function.  In this apodized case, as well, only the first term would appear in this equation.)  The remaining terms, comprising the infinite series,  become small for small $\rho$ when $N$ is large.  Effectively, we can ignore these terms once they reach amplitude well below the threshold we care about.  For detail on the derivation of this expression, see \mycitet{Van03} and \mycitet{Cad08}.  We note that, for the purposes of optimizing $A(r)$, the first term is linear in $A(r)$ while subsequent terms are not.  In general, when optimizing an occulter, we only use the first term, and select sufficient petals so that the first perturbation term is not important.

\myeq{prop1} determines the electric field at the entrance aperture of the telescope, and so is valid for both pure and hybrid occulters.  For pure occulters, the field at the image plane is found via a Fourier transform; for a hybrid, we use the propagation integrals for one or more APLCs.  (See \mycitet{Cad08} for more detail.)

\subsection{Optimization} \label{subsec:opt}

The basic optimization to make a pure occulter is discussed in detail in \mycitet{Cad08}; we review it briefly.  The optimization may be written in the form:
\begin{eqnarray}
    \mathrm{Minimize:}\quad &&c \label{opt1}\\
    \mathrm{subject\ to:}&&\quad Re(E_{\mathrm{apod}}(\rho)) - c \leq 0 \nonumber \\
    && -Re(E_{\mathrm{apod}}(\rho)) - c \leq 0 \nonumber \\
    &&Im(E_{\mathrm{apod}}(\rho)) - c \leq 0 \nonumber \\
    && -Im(E_{\mathrm{apod}}(\rho)) - c \leq 0 \nonumber \\
    &&\forall \quad \rho \leq \rho_{\mathrm{max}}, \quad \lambda \in [\lambda_{\mathrm{min}}, \lambda_{\mathrm{max}}] \nonumber \\
    && A(r) = 1 \quad \forall \quad 0 \leq r \leq a \nonumber \\
    && A'(r) \leq 0, \quad |A''(r)| \leq \sigma \quad \forall \quad 0 \leq r \leq R \nonumber
\end{eqnarray}
Each of these variables can be discretized and placed in a linear optimization, which produces a globally optimal solution that satisfies the given constraints.   In this case, it places a $||\cdot||_\infty$-norm bound on the real and imaginary parts of the electric field at the telescope aperture.  This can be easily extended to placing bounds on the electric field in the image plane, since the Fourier transform is a linear operator. Making a hybrid is a similar extension, except we use the propagation integrals for one or more APLCs, which are also linear operators.

One of the difficulties with the above approach is that, if the spectral band is wide, an occulter that provides sufficient suppression of the starlight has to be quite large and placed very far from the telescope.  An approach to mitigating that problem and designing smaller occulters, that we employ below, is to leverage the invariance of the field expression in \myeq{prop1} to the product $\lambda z$.   That is, we can scale both the wavelength and the distance by the same constant and achieve the identical shadow,
\begin{equation} \label{lambdaZ1}
    \lambda z = (\lambda c)\frac{z}{c}
\end{equation}
We exploit this fact by narrowing the design bandwidth at a given distance, and then moving the occulter closer for observations at longer wavelengths.  In other words, let the full spectral band over which we wish to investigate be $[\lambda_L, \lambda_H]$.  We create an optimization over a smaller band, $[\lambda_L, \lambda_M]$; with the spectral constraints reduced, the optimization can produce a smaller, closer occulter.  We then move the occulter inward from $z$ to $z/c$, where $c = \lambda_H/\lambda_M$.  The occulter now works over $[\lambda_L c, \lambda_H]$, and we can observe in the remainder of the band. Note that we must have $\lambda_L \lambda_H \leq \lambda_M^2$ to maintain continuity.  (If necessary, this could be repeated multiple times, for smaller band sections; the caveat is that the observation time is likely to increase, as the observations across the spectrum must be done in series rather than in parallel.)  The downside of this approach is that in the shifted band, the angular size of the occulter is larger by $c$, and so planets observable at the lower bandwidth may be attenuated or blocked by the occulter.

For a pure occulter, the effects are straightforward when the occulter is moved in---the PSF at $\lambda$ is now at $\lambda c$, identical except for an expansion in real units in the image plane.  (When optimizing for a pure occulter, the core of the PSF should be constrained slightly inward of the IWA in the image plane to compensate for this stretching.)  For a hybrid occulter, the results are more complicated: in general, we optimize the hybrid with a specific APLC, and we need to replace this with one designed for the longer wavelengths when the occulter is moved inward.  The PSFs then need to be verified to ensure that they perform as expected.

There may also be number of engineering constraints a design should meet.  Three of the most important, from the perspective of engineering the petals, are the width of the gap between petals, width of the tips of the petals, and length of the petals.  Petal length is included in \myeq{opt1} as $R-a$; the other two may be added by including the following constraints:
\begin{eqnarray}
    r A(r) \frac{2 \pi}{N} \geq \sigma_1 \label{con1} \\
    r (1-A(r)) \frac{2 \pi}{N} \geq \sigma_2 \label{con2}
\end{eqnarray}
where $\sigma_1$ is the minimum width for petal tips, and $\sigma_2$ is the minimum width for gaps between petals.

\subsection{Performance metrics}

The ultimate goal of an occulter or a hybrid is to reduce the light from the star to a point where it can be distinguished from the planet.  There are a few figures of merit for measuring how well this is being performed:
\begin{itemize}
\item Q-factor: the ratio between the peak of the planet signal and the star signal at the same location in the image plane.  We note that for this, it is important to consider the envelope of each PSF, as the PSFs will generally have valleys that should not be counted upon to measure the signal.
\item Suppression at the image plane: The peak intensity of the electric fields at the image plane at the location of the planet peak, at any wavelength.  For an occulter alone, this is generally Q-factor $\times$ a constant, as the occulter doesn't affect planets beyond its geometric IWA.  If you are looking at planets closer than the geometric IWA, or there is a coronagraph in the system, then the planet will be attenuated as well, and the Q-factor must be computed separately.  The image plane suppression should be chosen to maintain a high Q-factor after the error budget is taken into consideration.
\item Suppression at the aperture: the peak intensity of the electric fields across the telescope aperture at any wavelength.  This is a measure of how much of the light is being taken out of the system by the occulter.  For occulters alone, the image plane suppression is 10 to 100 times the aperture suppression; for a hybrid, it can vary considerably, if the coronagraph is expected to do a good deal of the suppression.
\end{itemize}

For THEIA, the occulter was designed to have a Q-factor of approximately 100 in the absence of aberrations and errors, in order to have sufficient signal to detect the planet in the presence of aberrations and other errors.  Since we are looking for Earth-like planets, the planet is assumed to have an intensity of $10^{10}$ at the aperture.  \mycitep{Des02}  While it didn't become an issue for THEIA, it is important to keep in mind the suppression at the image plane, and not just the Q-factor, as suppressing both star and planet significantly will increase the required integration times to detect the planet.

The inner working angle (IWA) of a hybrid system, the innermost angular separation at which we can detect a planet, depends both on the occulter and the coronagraph.  Its determination incorporates a number of factors:
\begin{itemize}
\item Geometric inner working angle: this is the angular size of the occulter in the sky, center to petal tip, and is equal to:
\begin{equation}
    \mathrm{IWA}_g = \sin{\left(\frac{R}{z}\right)} \approx \frac{R}{z}
\end{equation}
for small angles.  Beyond this point, we treat any planets as if they are unaffected by the occulter.  The geometric IWA is \emph{independent} of the aperture of the telescope; an occulter can have an almost arbitrarily small geometric IWA by adjusting the ratio of $R$ and $z$, and reoptimizing.
\item Mask size: The angular size of the mask in the APLC, if one is used.  We generally choose the mask size to be approximately equal to the geometric IWA.  Larger masks will attenuate the planets that the occulter did not; smaller masks are less effective at removing light from the system.
\item Half-power point: If a throughput curve of the entire system is made as a function of angular size, this is the angle at which the throughput drops to 1/2.  While this number requires a full system simulation, which makes it less amenable to rule-of-thumb design than the two above, it is also a much more accurate representation of the light attenuation.  The half-power point is generally smaller than the greater of the above two; if they're set to be equal, either one can be considered a conservative estimate for the half-power point.
\end{itemize}

For THEIA, we used $75$mas as the geometric IWA and mask size in our designs.  This number was chosen as a compromise between technical limitations---at $1000$nm, $75$mas is 1.45 $\lambda/D$, which means the coronagraph will barely do anything---and scientific goals, which would prefer the smallest angular size possible.

\subsection{Results and analysis} \label{subsec:ora}

For THEIA, we investigated four basic architectures:
\begin{enumerate}
\item An occulter alone, at a single distance from the telescope, providing all the suppression between $250-1000$nm.
\item An occulter alone, operating at two distances.  A different spectral band would be investigated at each distance, and the two together would span $250-1000$nm.
\item An occulter and coronagraph(s), operating at a single distance.
\item An occulter and coronagraph(s), operating at two distances.
\end{enumerate}

For the coronagraphs, we decided to use two apodized pupil Lyot coronagraphs (APLCs) \mycitep{Sou03}, one designed for a wavelength of $575$nm, and the other for a wavelength of $925$nm  (see \mytable{t2}).  Each coronagraph was more effective over its band than a single coronagraph attempting to cover the entire band.  Additionally, we decided that all of the suppression in the $250$-$400$nm band would be done with the occulter, to limit attenuation in the UV by the optics. The wavelength separating the two bands was pushed down to $700$nm from $750$nm as it was easier to design the occulter to work over the smaller band; the coronagraphs are identical between the two options.

For the multiple-distance occulter, the following additional constraints were applied:
\begin{itemize}
\item Petals shorter than $14$m (to ensure they fit in the fairing of existing launch vehicles)
\item Gaps between petals $\geq 1$mm (to ensure they can be manufactured)
\item Petal tip widths $\geq 1$mm (to ensure they can be manufactured)
\end{itemize}
No feasible solution meeting all of these constraints could be found for the single-distance occulter.

The most important result that came out of the optimizations was that, for the set of wavelengths under investigation, a hybrid adds little.  For a single-distance occulter, the occulter has to provide all of the suppression in the UV and in the near IR (a $75$mas mask is $1.45 \lambda/D$ in radius at $1000$nm---the coronagraph is doing very little).  In the two-distance arrangement, the results are similar---the occulter has to provide most of the light suppression at $1000$nm for a $75$mas mask.  Sec.\ref{subsec:why} provides a more in-depth discussion of these findings.  A $118$mas mask would be effective for planets beyond $118$mas, but would cut out any below this, removing much of the system's ability to characterize planets.  (An occulter with an angular size of $118$mas at the tips generally does not fully attenuate light at $75$mas, and so a planet could still be detected even when the occulter is closer in.)

\begin{figure}
\begin{center}
\subfigure{\includegraphics[width=3.5in]{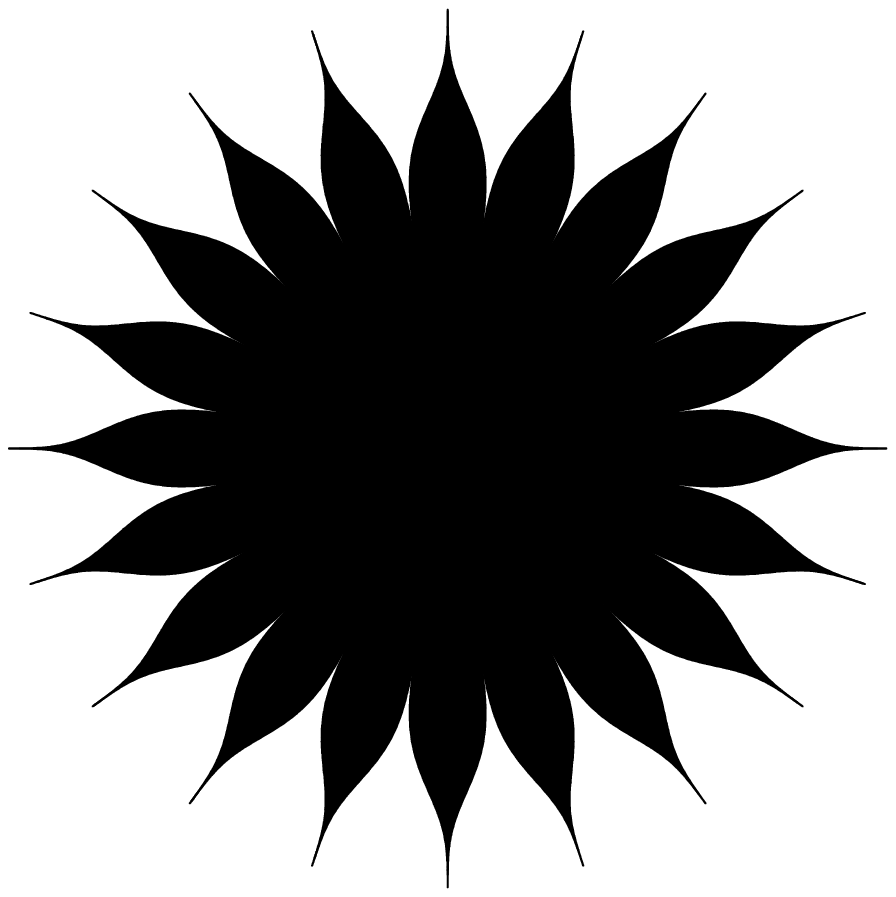}
\includegraphics[width=3.5in]{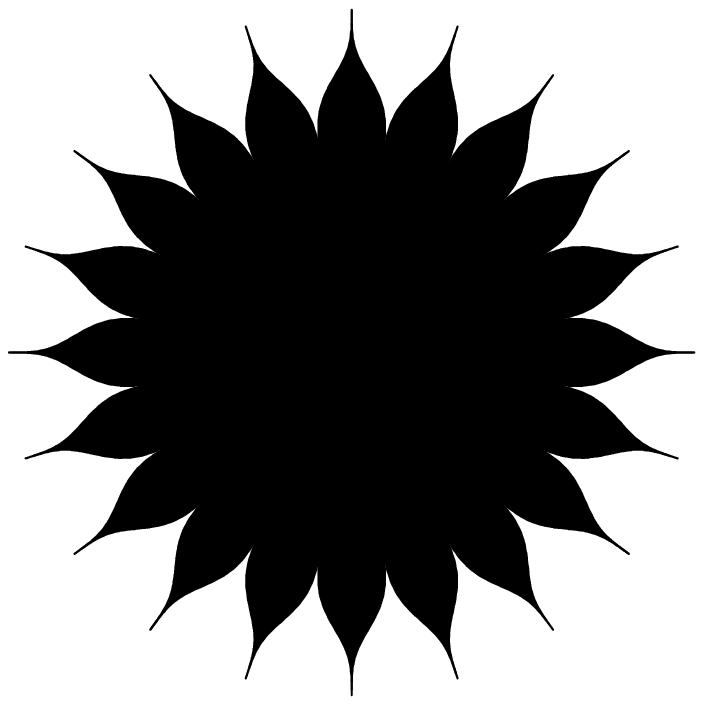}}
\caption{The two occulter designs, scaled proportionately.  Note that details, like gaps between petals, may be too small to see.  \emph{Left.} Single distance.  \emph{Right.} Multiple distance.}
\label{fig:occpix}
\end{center}
\end{figure}

\begin{figure}
\begin{center}
\includegraphics[width=6.25in]{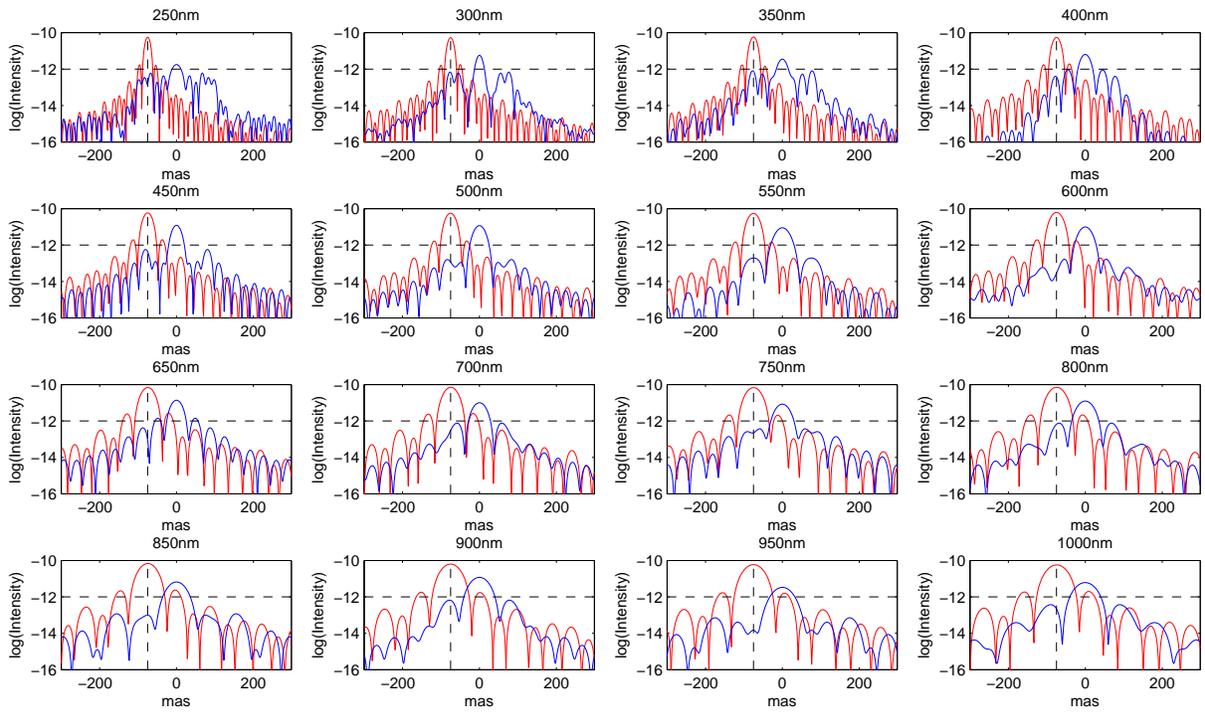} 
\caption{A series of slices through the PSF of the single-distance occulter at different wavelengths.  Star signal is in blue, planet is in red.}
\label{psfO1}
\end{center}
\end{figure}

In the end, a hybrid which meets the performance specification---starlight $\sim100\times$ dimmer than the planet at the planet location---turns out to be the same size and distance(s) from the telescope as an occulter alone.  The performance is slightly improved in the hybrid, in terms of star/planet ratio, but not significantly, and the already-faint planet light is attenuated further.  As a result, the top design for THEIA proved to be the multiple-distance occulter, with no coronagraph.  The designs are summarized in \mytable{t}.  Images of these occulters are shown in \myfig{fig:occpix}, and their PSFs in \myfigs{psfO1}{psfH2}.  Additionally, full-mission performance evaluations were conducted on the single-distance and multiple-distance occulters; these are summarized in Sec. \ref{sec:drm}.

\begin{figure}
\begin{center}
\subfigure{
\includegraphics[width=3.5in]{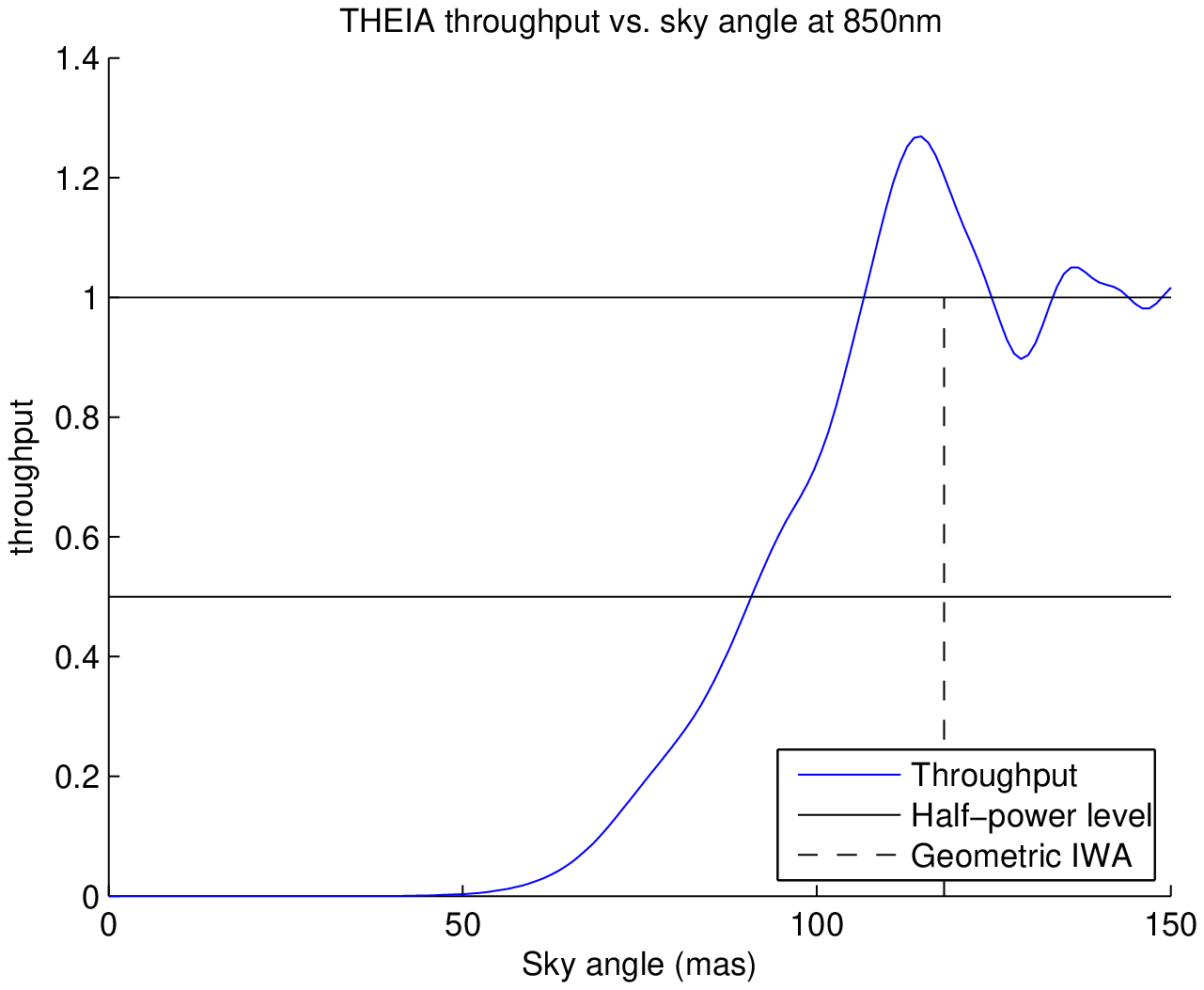}
\includegraphics[width=3.5in]{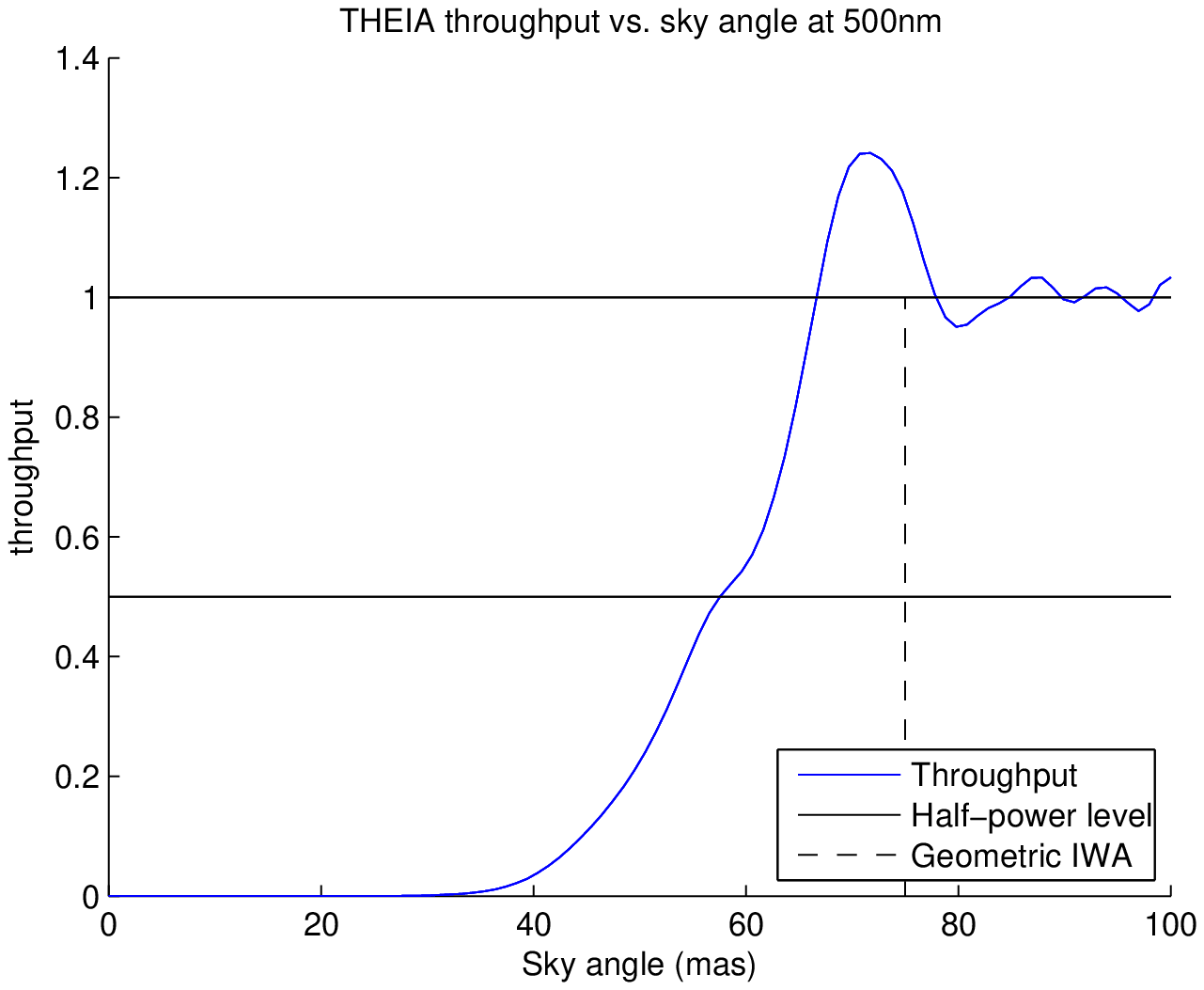}}
\caption{Throughput curves for the multiple-distance THEIA occulter.  \emph{Left.} Throughput at 850nm, occulter at 35000km. \emph{Right.} Throughput at 500nm, occulter at 55000km.}
\label{thNearFar}
\end{center}
\end{figure}

Throughput curves for two wavelengths at the near and far distances for the multiple-distance occulter can be seen in \myfig{thNearFar}.  One thing to note is that the throughput at certain angles is greater than one; this is caused by the occulter diffracting more light into the pupil of the telescope than would otherwise be present from a planet alone.   These structures persist even after radial averaging, and the magnitude of the fluctuations is large enough to affect the detection of planets near the geometric IWA  \mycitep{Sav09a}.

\section{THEIA System Design}\label{sec:design}

Our final design for the THEIA system consists of a an observatory spacecraft containing a 4-m primary telescope located at the Sun-Earth L2 point and the 40-m multidistance occulter flying in formation.  The observatory contains three instrument suites---the eXoplanet Characterizer (XPC) for planet detection and characterization, the Star Formation Camera (SFC) for wide-field UV/Visible astronomy, and the Ultraviolet Spectrograph (UVS) for high-resultion spectroscopy in the 100-300 nm wavelength range.  XPC consists of three science cameras operating in the UV (250-400 nm), Visible (400-750 nm), and Near-IR (750-1000 nm) plus two Integral Field Spectrometers (IFSs).  XPC also contains a fine occulter tracking camera (FOTC) for stationkeeping.  They are all fed by a 0.1 deg off-axis beam and picked off just before the Cassegrain focus of the telescope.

\begin{figure}
\begin{center}
\includegraphics[width=0.8\textwidth]{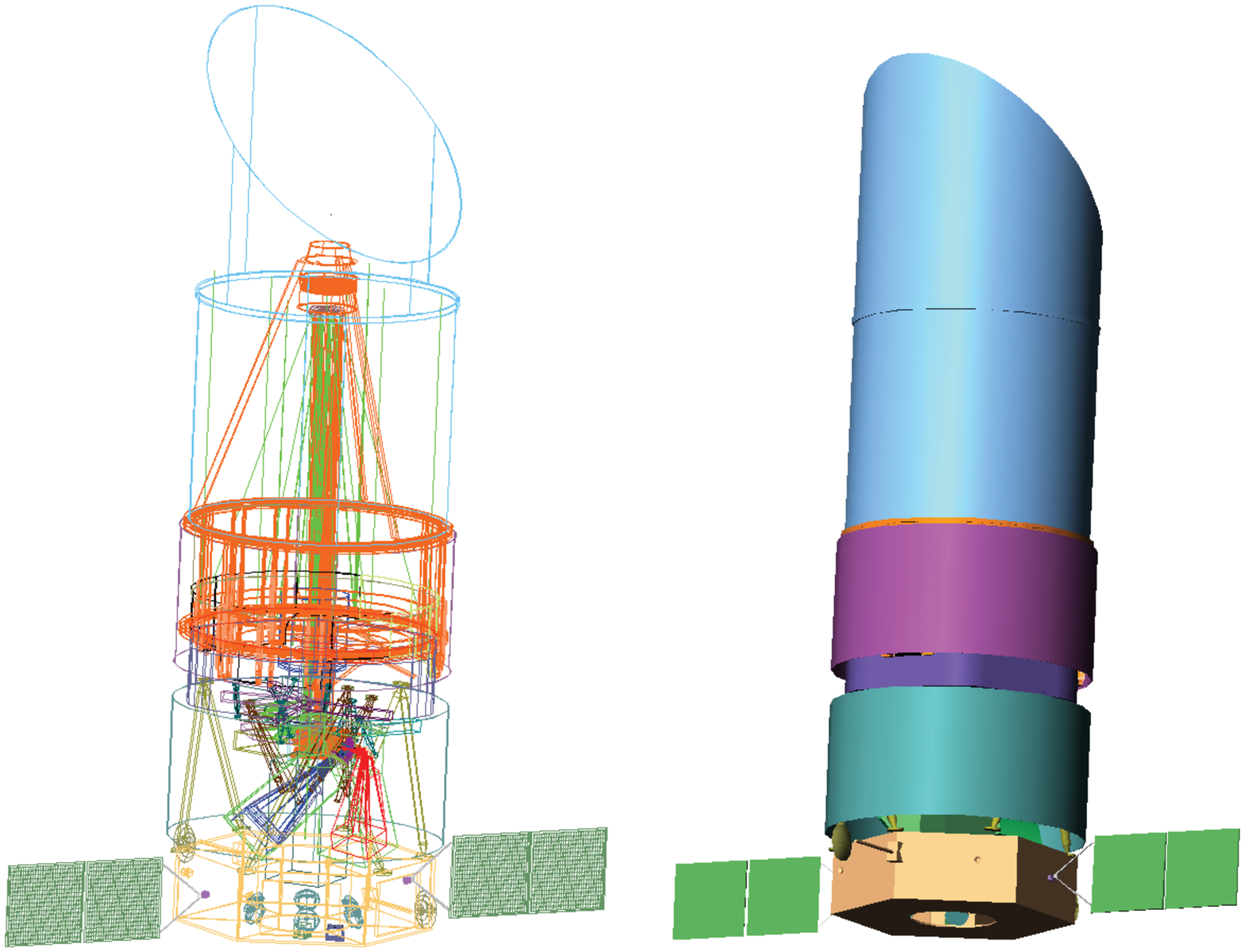}
\caption{The THEIA Observatory}
\label{fig:observatory}
\end{center}
\end{figure}

The THEIA observatory (see \myfig{fig:observatory}) is a natural evolution from HST and JWST.  The 4-m, on-axis telescope, and attached instruments, is mated with a conventional spacecraft system.  The observatory is 3-axis pointed within 3 arcsec using reaction wheels with hydrazine thrusters for momentum desaturation and stationkeeping.  The telescope bore-sight is pointed to within 30 milliarcsec using active struts and the final beam is pointed to within 4 mas using fine steering mirrors (FSMs).  Sensing is accomplished through a combination of rate gyros, fine guidance sensors, and star trackers.  The observatory communicates with the occulter spacecraft and Earth using separate S-band links.  The observatory can downlink up to 6 TBytes of data per day with a 1 Gbps downlink rate and 12 TBytes of storage. The observatory has a maximum expected mass of 5,700 kg as compared to a launch mass capacity of 6,300 kg. Observatory solar arrays are conservatively sized for 5kW of power at end of life, providing large margins.

THEIA's telescope is an on-axis three mirror anastigmat (TMA) optical design with a 4 m aperture and diffraction limited performance to 300 nm.   The relatively slow f1.5 Primary Mirror (PM) is a stiff, lightweight, closed-back design constructed of Corning Ultra Low Expansion glass with an Al/MgF$_2$ coating. The segmented lightweight honeycomb core is joined to the lightweighted pocket-milled mirror facesheets using a low temperature fusion process.  The PM is polished and final ion beam figured to a rms wavefront error less than 15 nm, only slightly better than Hubble's primary mirror ($\sim 20$ nm).  The Secondary Mirror (SM) mounts to a hexapod actuator that enables calibration of pointing and focus, though, due to the precision composite metering structures, SM re-alignments will be very infrequent.  The secondary mirror is Al/LiF coated for high reflectivity in the Far-UV to maximize UVS performance.

\begin{figure}
\begin{center}
\includegraphics[width=0.8\textwidth]{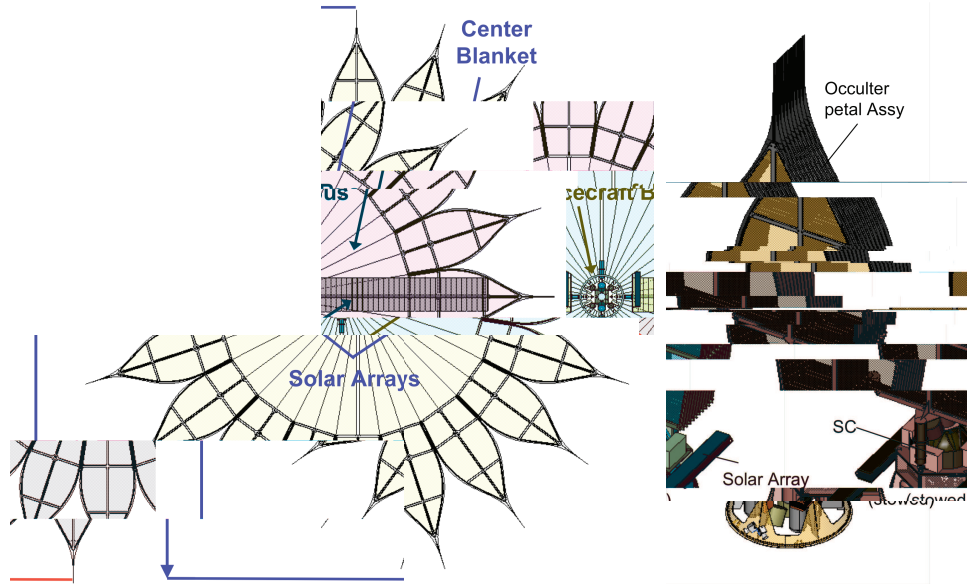}
\caption{The THEIA Occulter in deployed (\emph{Left}) and stowed (\emph{Right}) configurations.}
\label{fig:occulter_system}
\end{center}
\end{figure}

The occulter system consisting of the large starshade and attached spacecraft is shown in \myfig{fig:occulter_system}.
The starshade is comprised of of 2 major subsections, a 19.46 m diameter inner core composed of 3 layers of Kapton and an outer section of twenty 10.27 meter tall and 3.7 meter wide petals.  The precision shaped petal edge, defined for maximum light suppression, is machined into an extremely low CTE graphite epoxy sheet.  The sheet is bonded to a petal perimeter graphite epoxy box frame to provide structural support.  In the stowed configuration, the central core and petals mount to a graphite composite deployment deck, which in turn mounts to the spacecraft.  A central opening accommodates the recessed mounting of propulsive thrusters, antennas and laser beacons.   The stowed configuration fits inside a 5 m launch fairing with margin.  A truss structure supports the stowed petals and is jettisoned after launch.  Deployment is initiated by extending the entire stowed petal stack on two deployable booms.  This linear action unfolds the center-blanket assembly.  Once fully extended, the booms begin a rotation that initiates sequential deployment of the petals.  All petal hinge lines are controlled by redundantly actuated, passively damped, high accuracy hinges.  A simple sequencing cam between the primary hinge-lines of the petals also controls deployment.  As the booms rotate, the petals unfold.  The final position is controlled by stops designed into each hinge line.  Considerations of the hinge locations and gaps in the above optical analysis and design ensured that the needed starlight suppression was maintained.

The attached spacecraft provides power, propulsion, control, and communication.  The solar array is sized for 15 kW of end of life power to accommodate 2 NEXT-Ion thrusters firing simultaneously at maximum power (which can be adjusted in flight) plus 1 kW for other bus equipment.  This thruster subsystem consists of 6  thrusters in totalwith 3-for-2 redundancy on each side of the starshade.  During exoplanet observations, the occulter system is held on targets constrained to lie between 45 and 85 degrees from the sun line to avoid stellar leak into the telescope or reflections off the starshade.  Stationkeeping does not employ electric thrusters because they produce a bright plume potentially contaminating the observations.  The spacecraft is thus also equipped with a set of on/off hydrazine thrusters that control position to within $\pm 75$ cm.  A shutter is employed during the short, infrequent hydrazine pulses to avoid light contamination from the plume.  After a retargeting slew, the observatory/occulter formation is acquired in 4 overlapping stages of position sensing: 1) Conventional RF Ground tracking ($\pm 100$ km), 2) Observatory angle sensing of a Ka-Band beacon from the occulter spacecraft ($\pm 16$ km), 3) XPC IR imaging of a laser beacon ($\pm 70$ m) and 4) XPC IR imaging of light leaking around the occulter ($\pm 35$ cm).  The Observatory measures occulter range via the S-Band link. The occulter system has a maximum expected mass of 5,700 kg as compared to a launch mass capacity of 6,300 kg.  The remaining mass margin will be used for additional fuel for extended operations.

\section{Mission performance} \label{sec:drm}

An analysis of the performance of the single-distance and multiple-distance occulter designs was carried out in the context of a complete 5 year mission.  This work is presented in detail in \mycitet{Sav09a}, and is summarized here.

The mission parameters, including the telescope orbit, occulter propulsion system, telescope camera system and optics were all held constant between the two designs, with only the occulter throughput, occulter-telescope separation distance, and occulter dry mass varied.  The occulter dry mass becomes a significant parameter when assuming a maximum launch vehicle payload capacity, in which case the dry mass determines the total fuel carried by the occulter.  The mass for the multiple-distance occulter was determined via an extensive engineering assessment, and the single-distance occulter mass was then calculated by assuming the same subsystem mass and
scaling the occulter structure mass proportionately to the occulter area (assuming constant thickness).  The result is that while the
smaller, multiple-distance occulter has ample fuel capacity for the selected launch vehicle to complete its primary 5 year mission, the
heavier single-distance occulter often runs out of fuel before the five years are up.

Three primary metrics were considered in evaluating the performance of the occulter designs: the number of total planet detections,  the
number of unique planets detected, and the number of full spectral characterizations (to S/N of 11 for R=70 over the whole 250-1000nm band).  The mission rules required only one complete spectrum for each unique planet, and in the case of the multiple-distance occulter, two halves of the spectrum acquired at different observational epochs were counted as a complete spectrum.  The target planetary population was taken to be Earth-twins on habitable zone orbits, and hundreds of full mission simulations were carried out to find the average design performance.

\begin{figure}
\begin{center}
\includegraphics[width=3.5in]{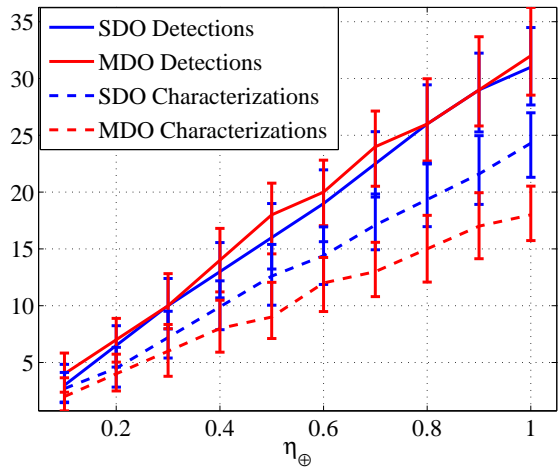}
\caption{Number of planets discovered and characterized spectrally over a 5-year mission, for single-distance and multiple-distance occulters, as a function of the frequency of terrestrial planets.}
\label{fig:DRM1}
\end{center}
\end{figure}

As reported in \mycitet{Sav09a}, in terms of total planet detections, the two designs yielded nearly identical results, with about 45 detections, on average, during a five year mission when assuming a universe with $\eta_\oplus = 1$ (one Earth-like planet per target star).  In terms of unique detections, the multiple-distance occulter marginally outperformed the single distance occulter, but only by about 5 additional unique detections on average in the $\eta_\oplus = 1$ case.  Both designs consistently found over 25 unique planets in these simulations.  These results can be seen in \myfig{fig:DRM1}.

The biggest difference was seen in the spectral characterizations, where the single-distance occulter outperformed the multiple-distance occulter by a factor of 1.5, with 25 full characterizations, on average.  Because the 20,000km slew that needs to be made by the occulter for a full spectral characterization could take up to 12 days to complete, few target stars had observing seasons long enough to accommodate both characterizations during one visit.  Even when the star remained observable for that long, the orientation of the planet's orbit would often cause the planet to become undetectable before the second half of the spectrum could be characterized. Furthermore, for the multiple-distance occulter, the 700-1000nm characterization occurred with a geometric working angle of 118mas, whereas the probability of detection of  a planet from the selected distribution peaks below 100mas, so that the higher wavelength characterizations needed to be longer than the lower wavelength ones in almost all cases.  This meant that even on subsequent visits where a previously found planet was re-detected, there would often not be enough time to carry out the entire higher wavelength characterization.

\begin{figure}
\begin{center}
\includegraphics[width=3.5in]{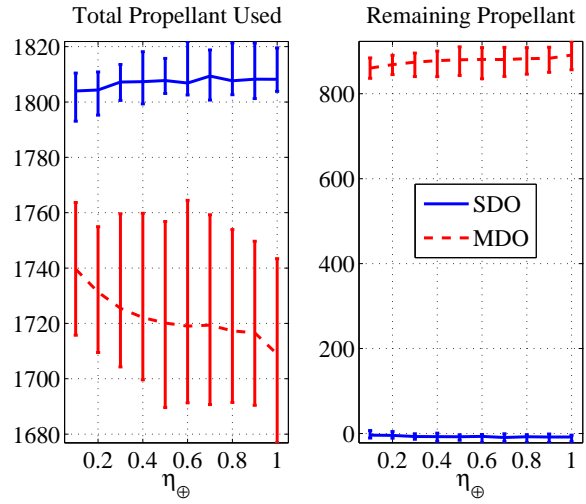}
\caption{Fuel use for single-distance and multiple-distance occulters after 5 years as a function of the frequency of terrestrial planets. \emph{Left.} Total fuel use over 5 years. \emph{Right.} Amount remaining after 5 years.  Under most scenarios, the single-distance occulter has used up its fuel before the 5 years are up.}
\label{fig:DRM2}
\end{center}
\end{figure}

Despite these factors, the multiple-distance occulter consistently completed the primary 5 year mission with enough fuel left over for at least two more years of operation, whereas the single-distance occulter consistently ran out of the fuel at the end of its primary mission.  This was due to higher amount of fuel required for slewing by the single distance occulter because of its larger separation from the telescope.  (See \myfig{fig:DRM2}.)

\section{Why were hybrid occulters infeasible?} \label{subsec:why}

\subsection{APLC/occulter hybrids}

The hybrid analysis for THEIA was based around APLCs, as they have certain advantages over other coronagraphs:
\begin{itemize}
\item Best star attenuation for a Lyot-type coronagraph \mycitep{Sou03}
\item Easy to design with an iterative algorithm \mycitep{Sou03}, \mycitep{Sou09}
\item Can be used with on-axis telescopes, including secondaries \mycitep{Sou09}
\item Can be used with arbitrary apertures \mycitep{Sou09}
\item Moderate insensitivity to small tilts and finite stellar size \mycitep{Guy06}
\item Moderate throughput \mycitep{Guy06}
\end{itemize}

The problems we found with these APLC hybrids are three-fold: First, the wavefronts from the star (and to a certain extent, the planet) can no longer be treated as flat and symmetric.  In particular, scattered light acquires a phase angle which would scatter light beyond the mask in the image plane of a Lyot-type coronagraph, which leaves it mostly unattenuated at the angular position of the planet.  Second, the hybrid has stricter alignment tolerances than an occulter alone, if the hybrid is doing a significant portion of the attenuation.  An occulter merely has to get the telescope in its shadow; a hybrid needs the telescope close to the center, or the asymmetry makes the performance poorer.

Third, the image-plane masks were sized to have their radius fall right at the geometric IWA of the occulter.  (The tilt of the planet wavefront is determined by geometry, rather than wavelength, and so the planet will appear at the same location in all bands.)  The APLC apodization is then determined using the central wavelength of the band.  Oversizing the mask would cut off the planet, while undersizing it would allow more starlight through the APLC.  Designing for the central wavelength was judged the best compromise: sizing the masks for the lowest wavelength in the band leaks too much starlight at the high end, and sizing the masks for the highest wavelengths attenuates the planet at the low end.  Additionally, the apodizer of the APLC will attenuate both star and planet light by a constant factor.  Thus, the hybrid attenuates the planet light as well as the star light, something the occulter does not do.

\begin{figure}
\begin{center}
\includegraphics[width=6.25in]{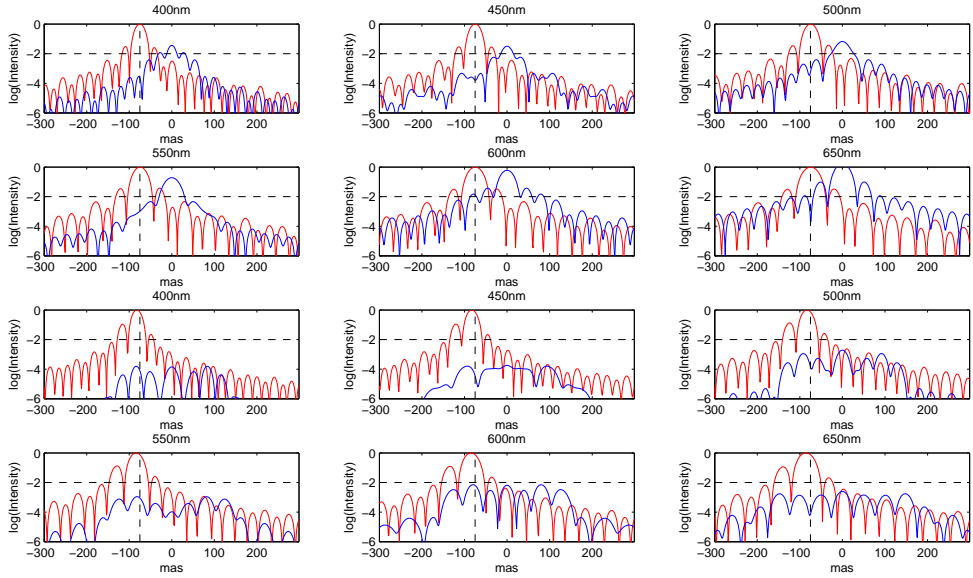}
\caption{This plot shows a comparison of the performance of an occulter and a hybrid system for THEIA.  The hybrid suppresses the cores greatly, but provides only a small suppression at the location of a $75$mas $10^{10}$ planet (vertical dashed line).  Each plot is scaled so that the peak of the planet in the image plane is at an intensity of 1, which makes the intensity of the star equal to $1/$Q-factor.  The star is in blue, the planet is in red.  \emph{Top 2 rows:} Occulter alone.  \emph{Bottom 2 rows:} Occulter and APLC.}
\label{fig:ovh}
\end{center}
\end{figure}

The hope in using hybrids was that the starlight would be attenuated enough relative to the planet light that the additional signal makes up for these drawbacks.  However, this turns out not to be the case.  Hybrid designs have the cores of their PSFs suppressed by two orders of magnitude or more, but the ratio between the intensity of the star and planet light at the location of the planet remains at a comparable level.  This can be seen in \myfig{fig:ovh}, which compares the performance of the multi-distance THEIA occulter with and without a coronagraph.  Essentially, there is no benefit to going to a hybrid system for the THEIA baseline scenario.

Introducing constraints on the magnitude of wavefront ripples in the optimization proved to be of limited use, as maintaining flatness across a wide spectral band tended to come at the expense of suppression at the aperture.  Even when it helped, it placed more severe lateral requirements on the placement of the occulter.  Optimizing the APLC apodizer and the occulter together proved of limited use as well, as the improvement in suppression was marginal and the requirements on lateral telescope-occulter alignment tightened; the optimization also became significantly more difficult.

\begin{figure}
\begin{center}
\includegraphics[width=3.5in]{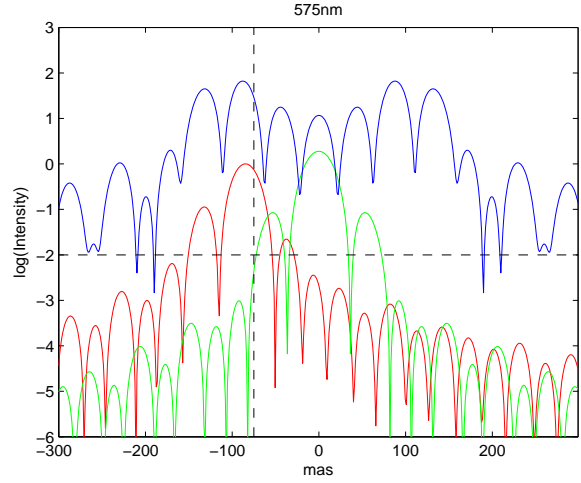}
\caption{This plot shows the difficulty of using a hybrid to provide suppression, comparing the performance of a hybrid system to an equivalent coronagraph with a plane wave incident. Each plot is scaled so that the peak of the planet in the image plane is at an intensity of 1, which makes the intensity of the star equal to $1/$Q-factor.  The star after the occulter is in blue, a flat wavefront of equivalent intensity is in green, and the planet is in red.}
\label{fig:hybridfail}
\end{center}
\end{figure}

This situation becomes worse when the hybrid is expected to take on some of the burden of suppressing the starlight.  In the THEIA case shown above, the occulter is doing all of the work suppressing the star light to $10^{12}$ in the image plane.  Consider \myfig{fig:hybridfail}, which shows a plot of the image plane through a hybrid system, with the planet peak normalized to intensity 1.  Here we have designed a system which explicitly splits the suppression between the occulter and coronagraph.  This system uses an occulter designed to provide $10^{6}$ or greater aperture suppression across a $400$nm-$700$nm band, and achieves the rest of the suppression with an APLC: APLC \#1 in \mytable{t}.  (The occulter here is $28.8$m across with $10$m petals, and has a geometric IWA of $75$mas.)

The blue line is the PSF from a small occulter designed to provide approximately $10^{6}$ aperture suppression; the remainder of the suppression ($\sim 10^6$) would be provided by the coronagraph.  The red line is the planet; the green line shows the performance of the system if the wave at the aperture had been a flat wavefront with intensity $10^{-6}$.  While a perfectly flat incident wavefront would have provided a Q-factor of approximately 100, an imperfect wavefront from the occulter with the same intensity has a Q-factor closer to $0.01$.  This decreased performance comes despite the use of flatness constraints.  This is one particular problem with hybrid systems---most coronagraphs assume a flat wavefront, and perform poorly otherwise.

Additionally, any design that wishes to detect the O$_3$ line needs to be able to detect down to $250$nm; in the UV, we cannot add a coronagraph, as too much light will be lost in the optics.  Thus, even if the occulter only needs to provide $10^6$ aperture suppression in the visible, the UV still needs on the order of $10^{10}$ aperture suppression.  This can be done in the optimization by weighting the constraints by wavelength, but the resulting occulters tend to be of a similar size to occulters that just provide $10^{10}$ suppression everywhere, so there is little advantage to asking the coronagraph to provide a lot of suppression when the telescope is going to be imaging in UV.

Another problem occurs in the NIR: a $75$mas mask at $1000$nm is $1.45 \lambda/D$ in radius.  For this small mask size, very little starlight can be removed by the coronagraph; the occulter still has to do most of the suppression.  Again, the optimization can tailor the aperture suppression levels by wavelength, but occulters that need to provide $10^{10}$ aperture suppression in the IR tend to be of similar size, even if less suppression is required elsewhere.

Some of these issues are specific to an APLC; a shaped pupil coronagraph, for example, could theoretically be designed without intermediate optics, and so could be used in the ultraviolet in a hybrid.  However, most coronagraphs require flat wavefronts to work effectively, and any hybrid design using a coronagraph to do a significant proportion of the suppression would be expected to also perform poorly without a wavefront control system to correct the incoming wavefront.  Introducing the wavefront control system would negate the advantages in cost and complexity gained by using an occulter.  Moreover, the wavefronts would not be corrected, merely distorted from their nominal shapes into plane waves, and the planet light would then be distorted as well.

\subsection{AIC/occulter hybrids}

The Achromatic Interfero-Coronagraph (AIC, \mycitep{Gay96, Bau00}) is a notable exception to this: the main advantage of the AIC that sets it apart from other coronagraphs when hybridizing with the occulter is that it does not require flat or achromatic wavefronts, so that designing the occulter to provide a flat illumination of the telescope is not necessary.  The AIC is a coronagraph based on interfering the pupil plane field $E_{\mathrm{pup}}$ with a flipped and phase-reversed copy of itself, resulting in a new, nulled pupil plane $[E_{\mathrm{pup}}(x, y) - E_{\mathrm{pup}}(-x, -y)]/2$.  The only requirement is that the field at the pupil plane of the telescope obey the symmetry relation $E_{\mathrm{pup}}(x, y) = E_{\mathrm{pup}}(-x ,-y)$, which essentially only requires that the telescope be in the center of the shadow (or that the shadow be uniform) and pointed well on the star.

An ideal AIC achieves total suppression of a point source while maintaining $50\%$ throughput of a planet $0.5 \lambda/$D away (and greater).  Another advantage of the AIC as compared to other coronagraphs is that the inner working angle of $0.5 \lambda/$D enables reduction in telescope size if telescope size is driven by the requirement that the IWA of the occulter and internal coronagraph be matched at the longest wavelength. Alternatively, for a given telescope size, occulter design can be chosen that provides a better inner working angle.  However, the main disadvantages of the AIC is its sensitivity to stellar angular size \mycitep{Guy06} and errors in occulter shape and alignment, and the fact that the planet throughput is reduced to $50\%$, split into two images of $25\%$ each.

\begin{figure}
\begin{center}
\includegraphics[width=3in]{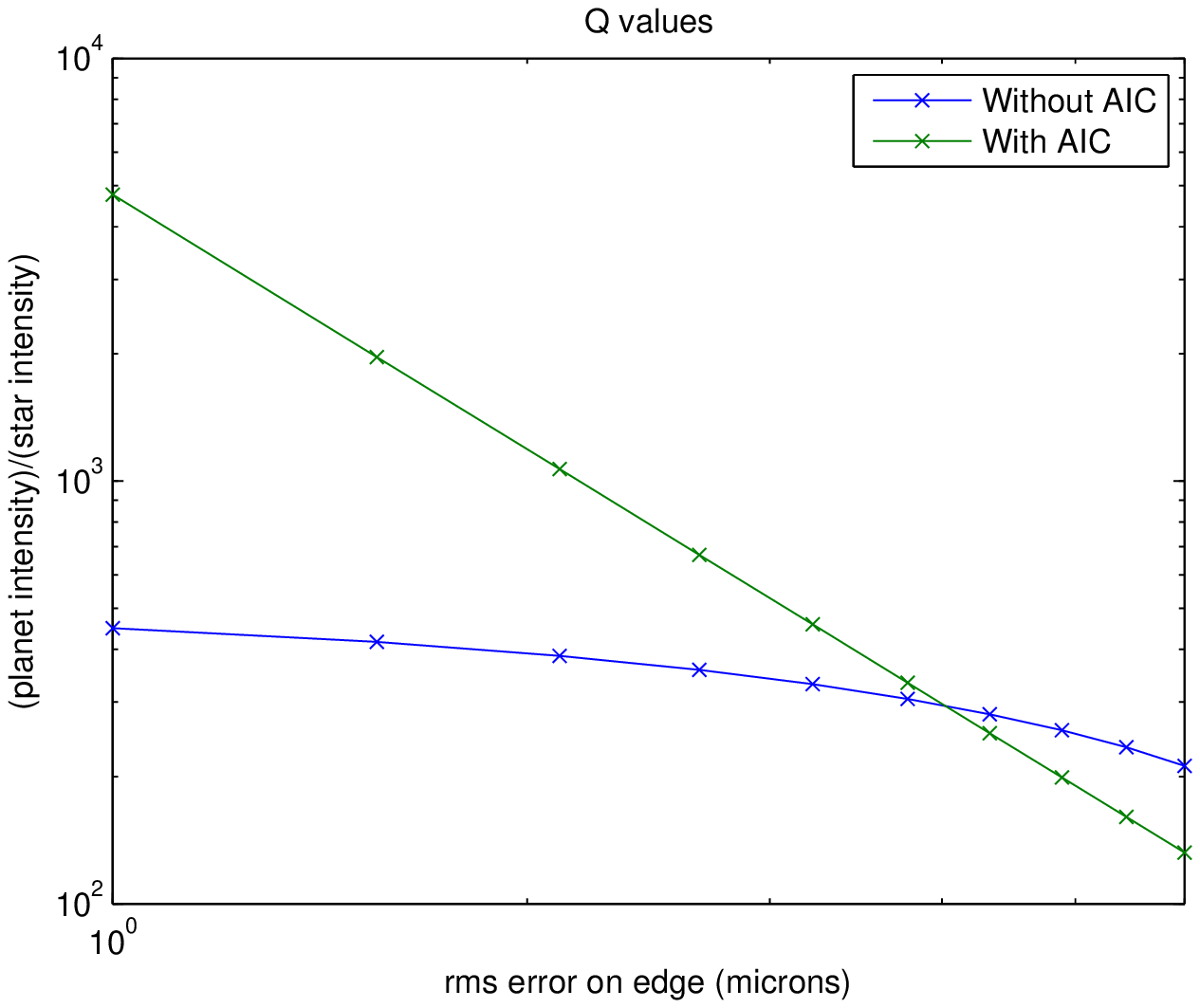}
\includegraphics[width=3in]{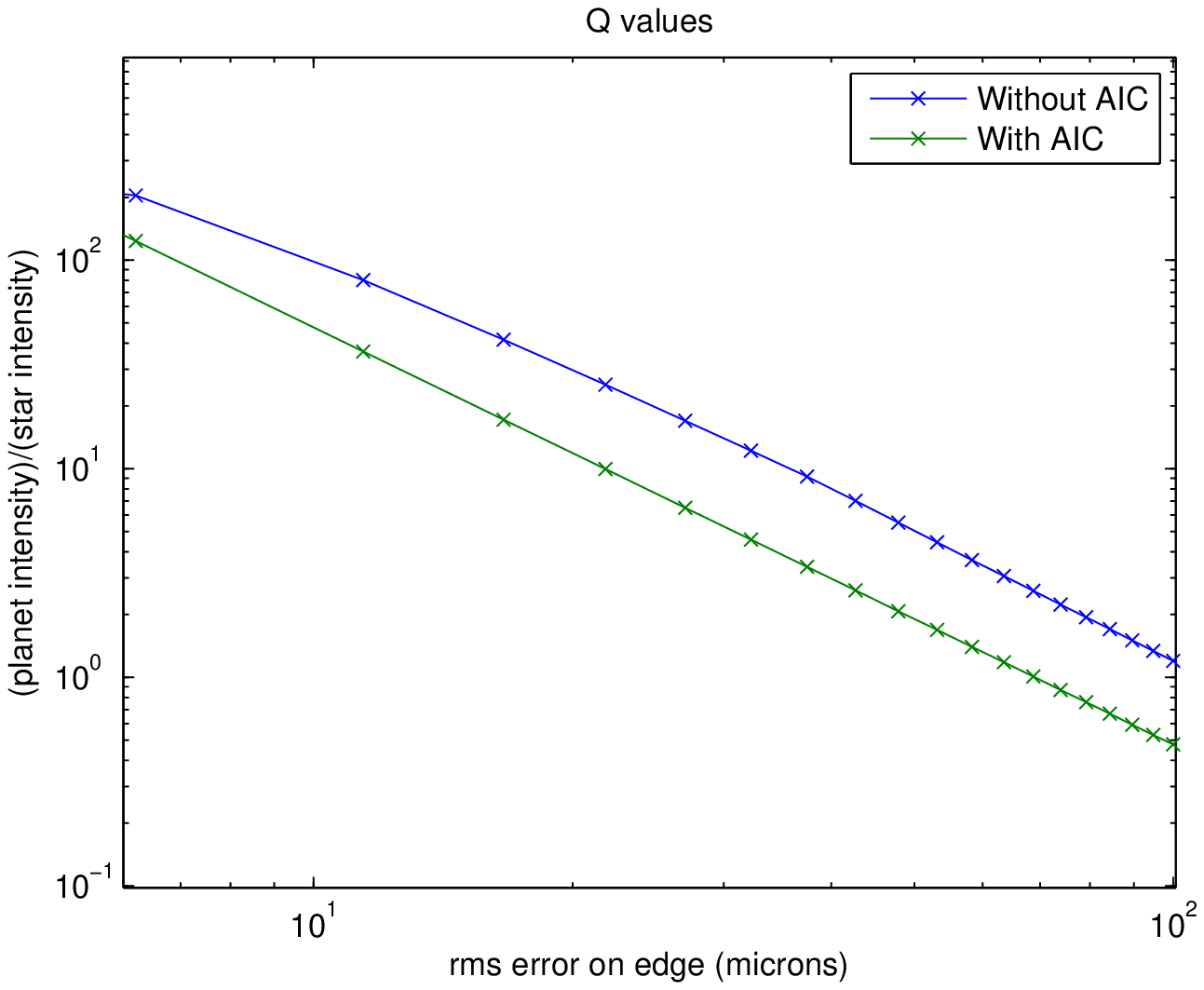}
\caption{Comparison of the performance of THEIA with and without an AIC, in the presence of random errors along the edges.  X-axis is rms error in deviation from the expected shape, y-axis is the ratio of planet light to star light (Q) at the inner working angle in the image plane. \emph{Left.}  Performance between 1 and 6 microns.  \emph{Right.} Performance between 6 and 100 microns.}
\label{fig:rmsErr}
\end{center}
\end{figure}

\myfig{fig:rmsErr} shows one example of this, comparing the ratio of the planet light and the star light in the presence of random edge errors.  When edge errors with size $3-4\mu$m rms or larger are applied to all petals, the occulter with AIC does no better than the occulter alone---in fact, worse, due in part to the $50\%$ throughput loss from the AIC.  This does not compare favorably to the $100\mu$m baseline edge error specified for THEIA.  (See \mycitet{Spe09} for a more complete list of allowable errors.)  It is worth noting that, in the presence of very small errors, using the AIC can improve performance, but at the expense of significantly tighter tolerances.  For this reason, it was not used for THEIA.

\section{Conclusions}

We present the results of our THEIA design study, with our best designs from four architecture comparisons.  Hybrid occulters were found to have a number of practical difficulties that limited their performance.  APLCs, and other architectures that relied on flat wavefronts, proved completely infeasible; while AICs worked, they required tighter tolerances than occulters alone, offsetting their potential advantages.  The best design would be able to image Earth-like planets around nearby stars with angular separations of $75$mas or less, and acquire spectra from $250$-$1000$nm.  This design includes constraints on the sizes of the gaps between petals; on the width of the petal tips; and on the lengths of the petals, to improve manufacturability.

Along with the optical design come the requirements on how accurately the design must be manufactured and held in position.  This tolerancing analysis of our occulter design will be presented in a subsequent paper.

\section*{Acknowledgments}

This work was performed under NASA contract NNX08AL58G, as part of the Astrophysics Strategic Missions Concept Studies (ASMCS) series of exoplanet concept studies.  The authors would like to thank Wes Traub for useful discussions.

\bibliography{refs}

\begin{thebibliography}{25}
\expandafter\ifx\csname natexlab\endcsname\relax\def\natexlab#1{#1}\fi

\bibitem[{{Basri} {et~al.}(2005){Basri}, {Borucki}, \& {Koch}}]{Bas05}
{Basri}, G., {Borucki}, W.~J., \& {Koch}, D. 2005, New Astronomy Review, 49,
  478

\bibitem[{Baudoz {et~al.}(2000)Baudoz, Rabbia, \& J.Gay}]{Bau00}
Baudoz, P., Rabbia, Y., \& J.Gay. 2000, Astron. Astrophys. Suppl. Ser., 141,
  319

\bibitem[{Brueckner {et~al.}(1995)Brueckner, Howard, Koomen, Korendyke,
  Michels, Moses, Socker, Dere, Lamy, Llebaria, Bout, Schwenn, Simnett,
  Bedford, \& Eyles}]{Bru95}
Brueckner, G.~E., {et~al.} 1995, Solar Physics, 162, 357

\bibitem[{Burrows(2005)}]{Bur05}
Burrows, A. 2005, Nature, 433, 261

\bibitem[{Cady {et~al.}(2008)Cady, Pueyo, Soummer, \& Kasdin}]{Cad08}
Cady, E.~J., Pueyo, L., Soummer, R., \& Kasdin, N. 2008, in Proceedings of
  SPIE--Space Telescopes and Instrumentation I: Optical, Infrared, and
  Millimeter Wave 2008, Vol. 7010

\bibitem[{Cash(2006)}]{Cas06}
Cash, W. 2006, Nature, 442, 51

\bibitem[{Copi \& Starkman(2000)}]{Cop00}
Copi, C., \& Starkman, G. 2000, Astrophysical Journal, 532, 581

\bibitem[{{Cowan} {et~al.}(2009){Cowan}, {Agol}, {Meadows}, {Robinson},
  {Livengood}, {Deming}, {Lisse}, {A'Hearn}, {Wellnitz}, {Seager},
  {Charbonneau}, \& {the EPOXI Team}}]{Cow09}
{Cowan}, N.~B., {et~al.} 2009, The Astrophysical Journal, 700, 915

\bibitem[{{Des Marais} {et~al.}(2002){Des Marais}, Harwit, Jucks, Kasting, Lin,
  Lunine, Schneider, Seager, Traub, \& Woolf}]{Des02}
{Des Marais}, D., {et~al.} 2002, Astrobiology, 2, 153

\bibitem[{Evans(1948)}]{Eva48}
Evans, J. 1948, Journal of the Optical Society of America, 38, 1083

\bibitem[{{Ford} {et~al.}(2001){Ford}, {Seager}, \& {Turner}}]{For01}
{Ford}, E.~B., {Seager}, S., \& {Turner}, E.~L. 2001, Nature, 412, 885

\bibitem[{Gay \& Rabbia(1996)}]{Gay96}
Gay, J., \& Rabbia, Y. 1996, Comptes rendus de l'Académie des sciences,
  S\'{e}rie II, 322, 265

\bibitem[{Guyon {et~al.}(2006)Guyon, Pluzhnik, Kuchner, Collins, \&
  Ridgway}]{Guy06}
Guyon, O., Pluzhnik, E.~A., Kuchner, M.~J., Collins, B., \& Ridgway, S.~T.
  2006, The Astrophysical Journal Supplement Series, 167, 81

\bibitem[{Howard {et~al.}(2000)Howard, Moses, \& Socker}]{How00}
Howard, R., Moses, J., \& Socker, D. 2000, in Proceedings of
  SPIE--Instrumentation for UV/EUV Astronomy and Solar Missions, Vol. 4139,
  259--283

\bibitem[{Kasdin {et~al.}(2009)Kasdin, Atcheson, Beasley, Belikov, Blouke,
  Cady, Calzetti, Copi, Desch, Dumont, Ebbets, Egerman, Fullerton, Gallagher,
  Green, Guyon, Heap, Jansen, Jenkins, Kasting, Keski-Kuha, Kuchner, Lee,
  Lindler, Linfield, Lisman, Lyon, MacKenty, Malhotra, McCaughrean, Mathews,
  Mountain, Nikzad, O'Connell, Oegerle, Oey, Padgett, Parvin, Prochaska,
  Rhoads, Roberge, Saif, Savransky, Scowen, Seager, Seery, Sembach, Shaklan,
  Shull, Siegmund, Smith, Soummer, Spergel, Stahl, Starkman, Stern, Tenerelli,
  Traub, Trauger, Tumlinson, Turner, Vanderbei, Windhorst, Woodgate, \&
  Woodruff}]{Spe09}
Kasdin, N.~J., {et~al.} 2009, {T}{H}{E}{I}{A}: Telescope for Habitable
  Exoplanets and Interstellar/Intergalactic Astronomy,
  http://www.astro.princeton.edu/~dns/Theia/nas\_theia\_v14.pdf

\bibitem[{{Kasting} {et~al.}(1993){Kasting}, {Whitmire}, \& {Reynolds}}]{Kas93}
{Kasting}, J.~F., {Whitmire}, D.~P., \& {Reynolds}, R.~T. 1993, Icarus, 101,
  108

\bibitem[{{Pall{\'e}} {et~al.}(2008){Pall{\'e}}, {Ford}, {Seager},
  {Monta{\~n}{\'e}s-Rodr{\'{\i}}guez}, \& {Vazquez}}]{Pal08}
{Pall{\'e}}, E., {Ford}, E.~B., {Seager}, S.,
  {Monta{\~n}{\'e}s-Rodr{\'{\i}}guez}, P., \& {Vazquez}, M. 2008, The
  Astrophysical Journal, 676, 1319

\bibitem[{{Savransky} {et~al.}(2009){Savransky}, {Kasdin}, \& {Cady}}]{Sav09a}
{Savransky}, D., {Kasdin}, N.~J., \& {Cady}, E. 2009, ArXiv e-prints

\bibitem[{Schultz {et~al.}(2003)Schultz, Jordan, Kochte, Fraquelli, Bruhweiler,
  Hollis, Carpenter, Lyon, DiSanti, Miskey, Leitner, Burns, Starin, Rodrigue,
  Fadali, Skelton, Hart, Hamilton, \& Cheng}]{Sch03}
Schultz, A., {et~al.} 2003, in Proceedings of SPIE--High-Contrast Imaging for
  Exo-Planet Detection, Vol. 4860

\bibitem[{Soummer {et~al.}(2003)Soummer, Aime, \& Falloon}]{Sou03}
Soummer, R., Aime, C., \& Falloon, P. 2003, Astronomy \& Astrophysics, 397,
  1161

\bibitem[{Soummer {et~al.}(2009)Soummer, Pueyo, Ferrari, Aime,
  Sivaramakrishnan, \& Yaitskova}]{Sou09}
Soummer, R., Pueyo, L., Ferrari, A., Aime, C., Sivaramakrishnan, A., \&
  Yaitskova, N. 2009, The Astrophysical Journal, 695, 695

\bibitem[{Spitzer(1962)}]{Spi62}
Spitzer, L. 1962, American Scientist, 50, 473

\bibitem[{Vanderbei {et~al.}(2007)Vanderbei, Cady, \& Kasdin}]{Van07}
Vanderbei, R., Cady, E., \& Kasdin, N. 2007, The Astrophysical Journal, 665,
  794

\bibitem[{Vanderbei {et~al.}(2003)Vanderbei, Spergel, \& Kasdin}]{Van03}
Vanderbei, R., Spergel, D., \& Kasdin, N. 2003, Astrophysical Journal, 599, 686

\bibitem[{Viv\`{e}s {et~al.}(2007)Viv\`{e}s, Lamy, Venet, Levacher, \&
  Boit}]{Viv07}
Viv\`{e}s, S., Lamy, P., Venet, M., Levacher, P., \& Boit, J. 2007, in
  Proceedings of SPIE--Solar Physics and Space Weather Instrumentation {I}{I},
  Vol. 6689

\end{thebibliography}
\bibliographystyle{apj}

\begin{table}
\begin{center}
\caption{Specifications of occulters operated at one or two distances from the telescope.}
\begin{tabular}{ccc}
\hline
\hline
 & 1-Dist. Occulter & 2-Dist. Occulter \\
\hline
Occulter radius (m) & 25.6 & 20 \\
Occulter nominal distance (km) & 70400 & 55000 \\
Occulter moved-in distance (km) & $\ldots$ & 35000 \\
Occulter spectral band at nominal distance (nm) & 250-1000 & 250-700 \\
Occulter spectral band at moved-in distance (nm) & - & 700-1000 \\
Occulter nominal IWA (mas) & 75 & 75 \\
Occulter moved-in IWA (mas) & $\ldots$ & 118 \\
Telescope diameter (m) & 4 & 4 \\
Number of petals & 20 & 20\\
Petal length (m) & 19 & 10 \\
Minimum gap between petals (mm) & 0.12 & 1.0 \\
Minimum width of petal tip (mm) & 1.62 & 1.0 \\
\hline
\end{tabular}
\end{center}
\label{t}
\end{table}

\begin{table}
\begin{center}
\caption{Specifications of coronagraphs operated at one or two distances from the telescope.}
\begin{tabular}{ccc}
\hline
\hline
 & Single Distance & Two Distance \\
\hline
APLC \#1 IWA (mas)  & 75 & 75 \\
APLC \#1 band (nm)  & 400-750 & 400-700 \\
\hline
APLC \#2 IWA (mas) & 75 & 75\\
APLC \#2 band (nm) & 750-1000 & 700-1000 \\
\hline
\end{tabular}
\end{center}
\label{t2}
\end{table}

\begin{figure}
\begin{center}
\includegraphics[width=6.25in]{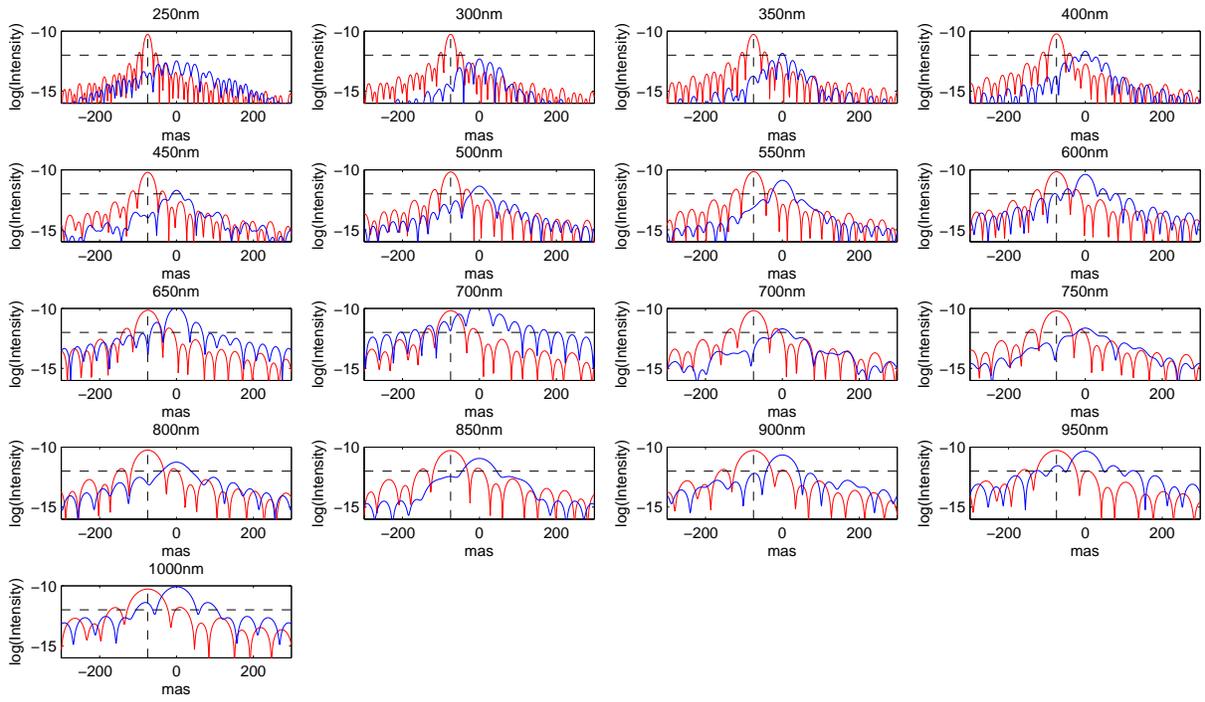}
\caption{A series of slices through the PSF of the multi-distance occulter at different wavelengths.  Star signal is in blue, planet is in red. \emph{250-700nm:} Outer distance.  \emph{700-1100nm:} Inner distance. }
\label{psfO2}
\end{center}
\end{figure}

\begin{figure}
\begin{center}
\includegraphics[width=6.25in]{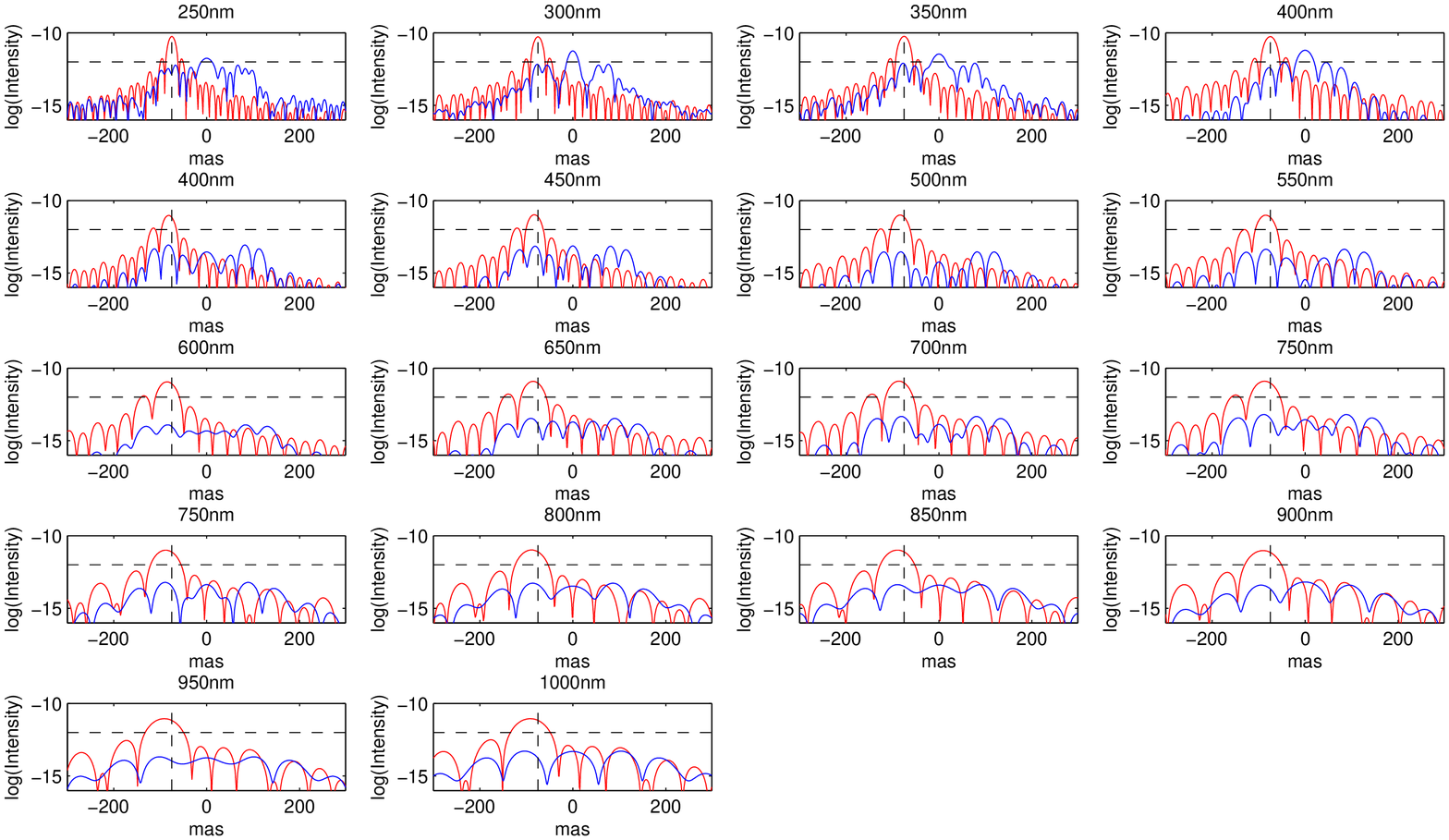}
\caption{A series of slices through the PSF of the single-distance hybrid at different wavelengths.  Star signal is in blue, planet is in red. \emph{250-400nm:} No coronagraph used.  \emph{400-750nm:} First APLC used.  \emph{750-1100nm:} Second APLC used.}
\label{psfH1}
\end{center}
\end{figure}

\begin{figure}
\begin{center}
\includegraphics[width=6.25in]{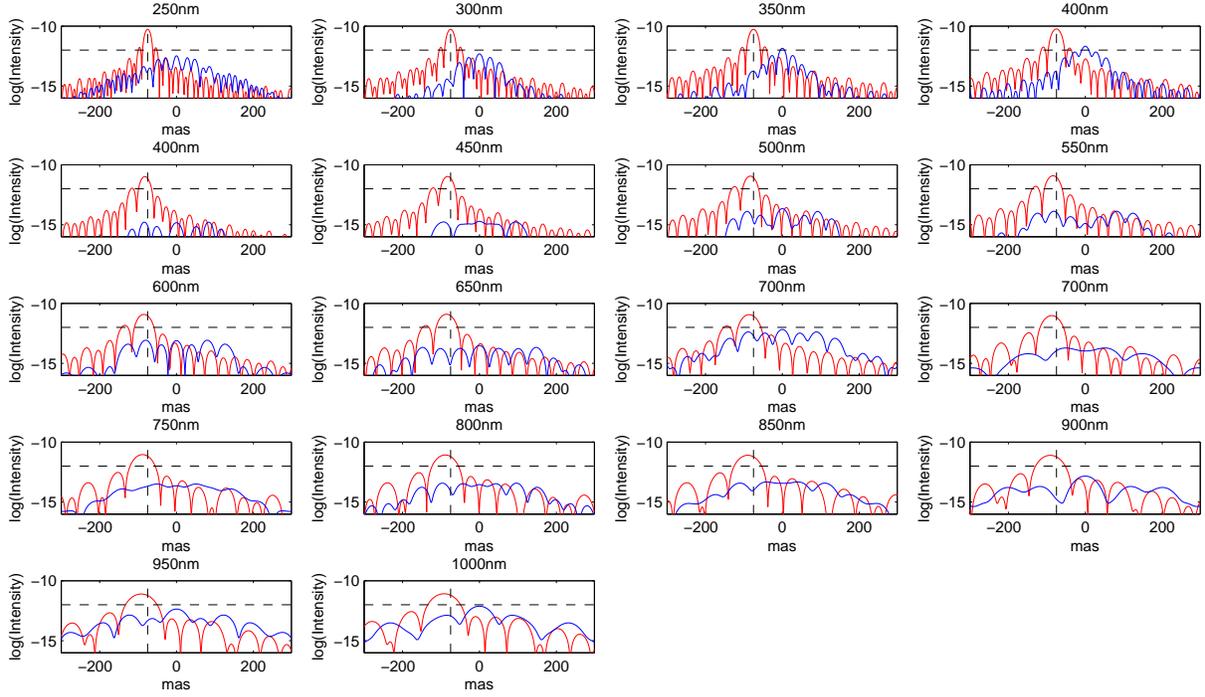}
\caption{A series of slices through the PSF of the multi-distance hybrid at different wavelengths.  Star signal is in blue, planet is in red. \emph{250-400nm:} No coronagraph used, outer distance.  \emph{400-700nm:} First APLC used, outer distance.  \emph{700-1100nm:} Second APLC used, inner distance.}
\label{psfH2}
\end{center}
\end{figure}

\end{document}